\documentclass[namedreferences]{solarphysics}
\usepackage[optionalrh,solaromanenum]{spr-sola-addons} 
\usepackage{graphicx}                    
\usepackage{color}                       
\usepackage{url}                         
\usepackage[pdfborder={0 0 0 },urlcolor=blue,breaklinks]{hyperref}

\newcommand{\kms}{km~s$^{-1}$}
\newcommand{\ion}[2]{#1\,{\sc #2}}

\newcommand{\ecss}{erg~cm$^{-2}$~s$^{-1}$~sr$^{-1}$} 
\newcommand{\hinode}{\emph{Hinode}}
\newcommand{\as}{${^\prime}{^\prime}$}
\newcommand{\solarsoft}{SolarSoft}

\begin{document}

\begin{article}

\begin{opening}

\title{\emph{Solar Dynamics Observatory} and \emph{Hinode} Observations of a Blowout
  Jet in a Coronal Hole}

%
\author{P.R.~\surname{Young}$^{1}$\sep
        K.~\surname{Muglach}$^{2,3}$
       }

%
\runningauthor{P.R. Young, K. Muglach}
\runningtitle{SDO and Hinode Observations of a Blowout Jet in a
  Coronal Hole}

%
  \institute{$^{1}$ College of Science, George Mason University, 4400
    University Drive, Fairfax VA 22030, USA\\
                     email: \href{mailto:pyoung9@gmu.edu}{\textsf{pyoung9@gmu.edu}} \\
             $^{2}$ Code 674, NASA Goddard Space Flight Center,
    Greenbelt, MD 20771, USA\\
                     email: \href{mailto:kmuglach@gmx.de}{\textsf{kmuglach@gmx.de}} \\ 
             $^{3}$ ARTEP, Inc., Ellicott City, MD 21042, USA
             }

\begin{abstract}
A blowout jet occurred within the south coronal hole on  9 February
2011 at 09:00~UT and was observed by the \emph{Atmospheric Imaging
Assembly} (AIA) and \emph{Helioseismic and Magnetic Imager} (HMI)
onboard the \emph{Solar Dynamics Observatory},  and the \emph{EUV
Imaging Spectrometer} (EIS) and \emph{X-Ray Telescope} (XRT) onboard
the \hinode\ spacecraft during coronal hole monitoring performed as
part of \hinode\ Operations Program No.~177. Images from AIA show
expanding hot and cold loops from a small bright point with plasma
ejected in a curtain up to 30~Mm wide. The initial intensity front of
the jet had a projected velocity of 200~\kms\  and line-of-sight (LOS)
velocities measured by EIS are between 100 and 250~\kms. The LOS
velocities increased along the jet, implying an acceleration mechanism
operating within the body of the jet. The jet plasma had a density of
$2.7\times 10^8$~cm$^{-3}$ and a temperature of 1.4~MK. During the
event a number of bright kernels were seen at the base of the bright
point. The kernels have sizes of $\approx$~1000~km, are variable in
brightness, and have lifetimes of 1\,--\,15~minutes. An XRT filter
	ratio yields temperatures of 1.5\,--\,3.0~MK for the kernels. The
	bright point existed for at least ten hours, but  disappeared within
	two hours after the jet, which lasted for 30~minutes. HMI data reveal
	converging photospheric flows at the location of the bright point, and
	the mixed-polarity magnetic flux canceled over a period of four hours
	on either side of the jet.
	\end{abstract}

	%

	\end{opening}

	\section{Introduction}

	Solar jets are recognized in image sequences as dynamic structures
	that yield a narrow, columnar structure extending away from the solar
	surface. Coronal holes present particularly favorable locations for
	studying coronal jets due to the low intensity of the background
	corona. Surveys of coronal hole jets  performed by
	\inlinecite{nistico09} and \inlinecite{moore10} have identified
	different types of jets and, in particular, one type that appears similar
	to a coronal mass ejection (CME) although on a much smaller spatial
scale.  An example of this type of event is presented here, uniquely
combining coronal spectroscopic data from the \emph{EUV Imaging Spectrometer}
(EIS: \opencite{culhane07})
onboard the \hinode\ spacecraft \cite{kosugi07},
and high-cadence, high-spatial resolution data from the \emph{Atmospheric
Imaging Assembly} (AIA: \opencite{lemen-aia}) and \emph{Helioseismic
and Magnetic Imager} (HMI: \opencite{scherrer-hmi}) onboard the
\emph{Solar Dynamics Observatory} (SDO). 

\inlinecite{nistico09} performed a survey of coronal hole jets observed
with the \emph{Sun Earth Connection Coronal and Heliospheric Investigation}
(SECCHI) instruments on the two \emph{Solar Terrestrial Relations
Observatory} (STEREO) spacecraft, and five of 
the 79 events were considered to be small versions of coronal mass
ejections dubbed ``micro-CMEs''. Such events were identified as
having a three-part structure of bright leading edge, dark void, and
bright trailing edge in analogy to the structures seen when CMEs are
viewed at the limb. Note that two of the five events demonstrated a
helical structure.

A survey of jets observed with the \emph{X-Ray Telescope} (XRT: \opencite{golub-xrt}) onboard the
\hinode\ spacecraft was performed by \inlinecite{moore10} who categorized
jets as ``standard'' or ``blowout''. The standard jets are the
equivalent of the categories ``lambda'' and ``Eiffel Tower'' used
by \inlinecite{nistico09}, while blowout jets are the equivalent of the
micro-CME jets of \inlinecite{nistico09}. \inlinecite{moore10} defined
a blowout jet to be one in which the base arch of the bright point is
blown open during the event, whereas the arch remains closed for
standard jets.
(The base
arch is an X-ray bright loop structure that spans the base of the
jet.) 
By inspecting cooler images from the EUVI instruments onboard the
STEREO spacecraft for some of the events, \inlinecite{moore10} concluded
that the blowout jets also exhibit the expulsion of cool material.

In the present article we will present photospheric magnetic field, and
coronal spectroscopic and imaging data for a blowout jet. The magnetic
evolution of jets has been previously discussed and a key result is
that convergence and cancellation of magnetic elements are commonly associated. For
example, \inlinecite{wang98} found that 80\,\%\ of H$\alpha$ jets in quiet
solar regions are associated with converging magnetic dipoles, while
\inlinecite{chifor08-jet2} found that an active region coronal jet was associated
with repeated flux cancellation. Two examples of coronal jets at a
coronal hole boundary were presented by \inlinecite{yang11}, both
identified with flux cancellation. A survey of the magnetic signatures
of X-ray brightenings within a coronal hole observed with the \emph{Solar
Optical Telescope} (SOT: \opencite{tsuneta-sot}) and XRT onboard \hinode\ was presented by
\inlinecite{huang12} who found that all of the 22 X-ray brightenings were
associated with canceling flux. One of these events was a jet. Flux
cancellation was also observed in many cases of transition region
explosive events by \inlinecite{muglach08}.  An
event described as a standard-to-blowout jet exhibited flux emergence,
convergence, and cancellation in observations presented by
\inlinecite{liu11}.

There have been several spectroscopic studies of coronal hole jets
with EIS, and we focus on those for which velocity measurements for
the plasma within the jet (rather than the bright point) were made, as
this is the measurement made in the present article. \inlinecite{kamio07}
found line-of-sight (LOS) blueshifts of about 
30~\kms\ in the \ion{He}{ii} 256.3\,\AA\ and \ion{Fe}{xii} 195.1\,\AA\
emission lines for one jet observed just inside the solar limb. A
coronal background subtraction however, such as performed in the
present article, may have led to higher velocities. A subsequent study by
\inlinecite{kamio09} did perform two Gaussian fits to two separate
jets observed in \ion{Fe}{xii} 195.1\,\AA, yielding LOS velocities of
$-50$ and $-96$~\kms\ for the jet plasma. A jet observed by
\inlinecite{doschek10} showed LOS velocities of $\approx -15$ to
$-20$~\kms, although this is again the Doppler shift of the combined
background plus jet emission line profile. \inlinecite{doschek10} also
derived temperatures of 1.1\,--\,1.3~MK in the jet and densities of 5\,--\,70
$\times$ $10^7$~cm$^{-3}$.

A detailed study of a bright jet in a low latitude coronal hole was
performed by \inlinecite{madjarska11}. LOS velocities within the broad jet
ranged from $-30$ to $-150$~\kms, although for a small region at one
side of the jet a clear two component structure to the emission lines
could be identified. Just above the bright point the velocity of this
component was $-310$~\kms, decreasing to $-150$~\kms\ over a projected
distance of about 10~Mm (measurements made in \ion{Fe}{xii}
195.1\,\AA). This high-velocity component was seen over a wide range of
temperatures, from 0.5~MK to 2.0~MK.

Section~\ref{sect.obs} summarizes the types of observations used in
the present article, and
Section~\ref{sect.overview} gives an overview of the
event. Section~\ref{sect.bp} discusses the evolution of the bright point
that gives rise to the jet, and Section~\ref{sect.jet} discusses the jet
itself. A summary of the results is given in
Section~\ref{sect.summary}. For the remainder of this article we will refer to the event as a blowout
jet. 


\section{Observations}\label{sect.obs}

The \hinode\ observations discussed here were obtained through \hinode\
Observing Program No.~177 (HOP 177), entitled ``Magnetic structures within
coronal holes'' (PI: P.R.~Young). 
Observations were obtained during
8\,--\,10 February 2011 when the south polar hole had an equatorial
extension, perhaps related to active region AR 11156, which lay at the
north boundary of the coronal hole.
A number of distinctive features were
identified in line-of-sight (LOS) velocity images obtained from the
\ion{Fe}{xii} 195.12\,\AA\ emission line, the majority of which were
identified to be jets. A summary of these jet observations will be
given in a future publication, and the present article focuses on one of the
jets, which is classed as a blowout jet. The major evolution of the
jet and associated bright point occurred between 08:45 and 09:20~UT on
February 9.

EIS  captured the jet with a raster that began at
09:05~UT on 9 February 2011. The raster scanned an area of 179\as\
$\times$ 512\as\ with a 60-second exposure time at each slit position,
giving a raster duration of 62~minutes. Many emission lines were observed
although this article principally focuses on \ion{Fe}{xii} 195.12\,\AA\
(formed at 1.5~MK). The EIS data were calibrated using the standard
options recommended in the EIS data-analysis
guide (\href{http://solarb.mssl.ucl.ac.uk:8080/eiswiki/Wiki.jsp?page=EISAnalysisGuide}{\textsf{solarb.mssl.ucl.ac.uk:8080/eiswiki/Wiki.jsp?page=EISAnalysisGuide}}),
and Gaussian fitting of the emission lines was performed using
software described by \inlinecite{eis_sw16}.

XRT obtained images in the Be-thin and Ti-poly filters at approximately 150-second
cadence, however there is a data gap between 09:03 and 09:25~UT. Note
that two Ti-poly exposures with different exposure times are taken for
each of the Be-thin exposures. The Be-thin filter has a greater
sensitivity to hot plasma than the Ti-poly filter.

The AIA and HMI  instruments onboard SDO
observe the entire Sun with regular, high temporal cadences and so both
instruments captured the evolution of the jet. AIA obtains images in
seven EUV filters at 12-second cadence, and two UV filters at 2-second
cadence. In this article we use the notation ``\emph{A171}'', ``\emph{A193}'', \emph{etc.}, to
refer to the AIA filters centered at 171 and 193\,\AA. 
HMI takes polarized full disk images with two camera systems: the
front camera measures the Stokes-$I$ and -$V$ parameters only and the
data pipeline yields LOS magnetograms, Dopplergrams and intensity
images at a 45-second cadence; the side camera measures all four Stokes
parameters to allow the full vector magnetic field to be computed at a
720-second cadence. The data pipeline for the side camera also yields LOS
magnetograms at a 720-second cadence, which have a somewhat higher
sensitivity than the front camera 45-second magnetograms \cite{liu12}. Both
types of LOS magnetograms are used in the present article.

Calibration software for all datasets was obtained from the IDL
\solarsoft\ library (\href{http://www.lmsal.com/solarsoft/}{\textsf{www.lmsal.com/solarsoft/}}).
For all images shown in this article, the listed time corresponds to the
midpoint of the exposure, given in  Coordinated Universal Time (UTC) format.
Three movies created from \emph{A193} data and one from HMI data are presented in the following sections.
Additional movies for the event, generated from  XRT, HMI, and other
AIA filters, 
are available at \href{http://pyoung.org/jets/hop177/jet\_27.html}{\textsf{pyoung.org/jets/hop177/jet\_27.html}}.

\section{Event Overview}\label{sect.overview}

\begin{figure} 
\centerline{\includegraphics[width=1.0\textwidth]{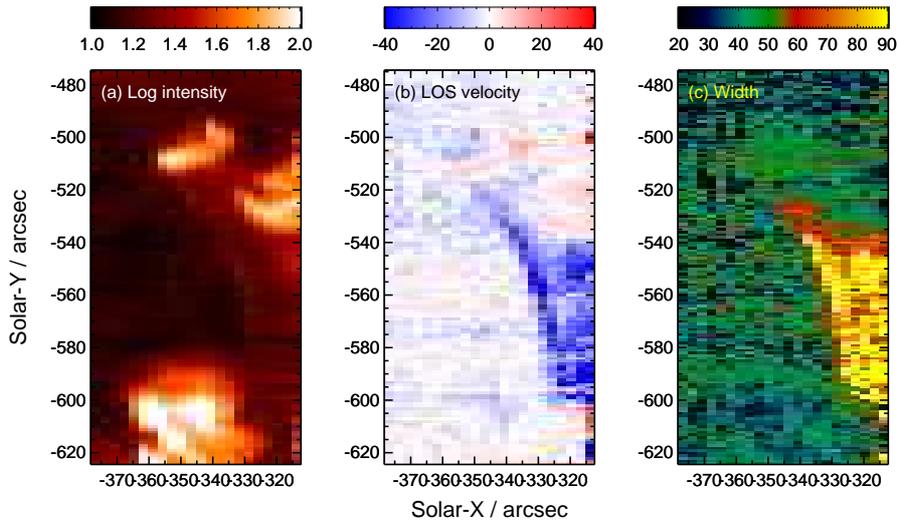}}
\caption{Images of the jet derived from single Gaussian fits to the
  \ion{Fe}{xii} 195.12\,\AA\ 
  emission line observed by EIS. The left panel shows the logarithm of
  intensity in units \ecss, the middle panel
  shows LOS velocities in \kms, and the right panel shows
  line width in m\AA. The instrumental line
  width has been removed.}
\label{fig.eis}
\end{figure}

Jets were identified in the coronal hole
by fitting a 
single Gaussian function to the \ion{Fe}{xii} 195.12\,\AA\ emission line
and constructing line-of-sight Doppler velocity maps from each EIS
raster. The blowout event presented here showed the largest Doppler
signatures of all the jets identified from the dataset.
Figure~\ref{fig.eis} shows the discovery image 
together with intensity and line width images. EIS
rasters from right to left and the images shown in
Figure~\ref{fig.eis} were obtained from 09:06 to 09:36~UT. The first eight
exposures were badly affected by particle hits during a South Atlantic
Anomaly encounter; thus the first useful exposure was at 09:14~UT. The
bright point at the base of the event was observed between 09:21 and 09:29~UT.

The intensity image from Figure~\ref{fig.eis} shows a faint, narrow
structure coming from the small bright point in the top-left of the
raster but otherwise is not distinctive. The velocity and line width
images, however, show a very clear structure: a thick column (observed
between 09:14 and 09:18~UT) with a short, narrow leg that connects to
the bright point. 

Movie 1 shows a sequence of \emph{A193} images at one-minute cadence with the
field-of-view cropped to show the jet and the bright point.
The movie covers the
period 08:30 to 09:39~UT, and it shows expanding loops coming from the
bright point in the top-left of Figure~\ref{fig.eis} over the period
08:50 to 08:59~UT with a wide jet becoming prominent between 09:00 and
09:23~UT. The jet expands laterally and fades over this time
period. The shape of the jet in the EIS velocity image
(Figure~\ref{fig.eis}b) matches the shape of the \emph{A193} structure. Note
that the EIS velocity signature is much more prominent than the \emph{A193}
intensity structure.

\begin{figure} 
\centerline{\includegraphics[width=1.0\textwidth]{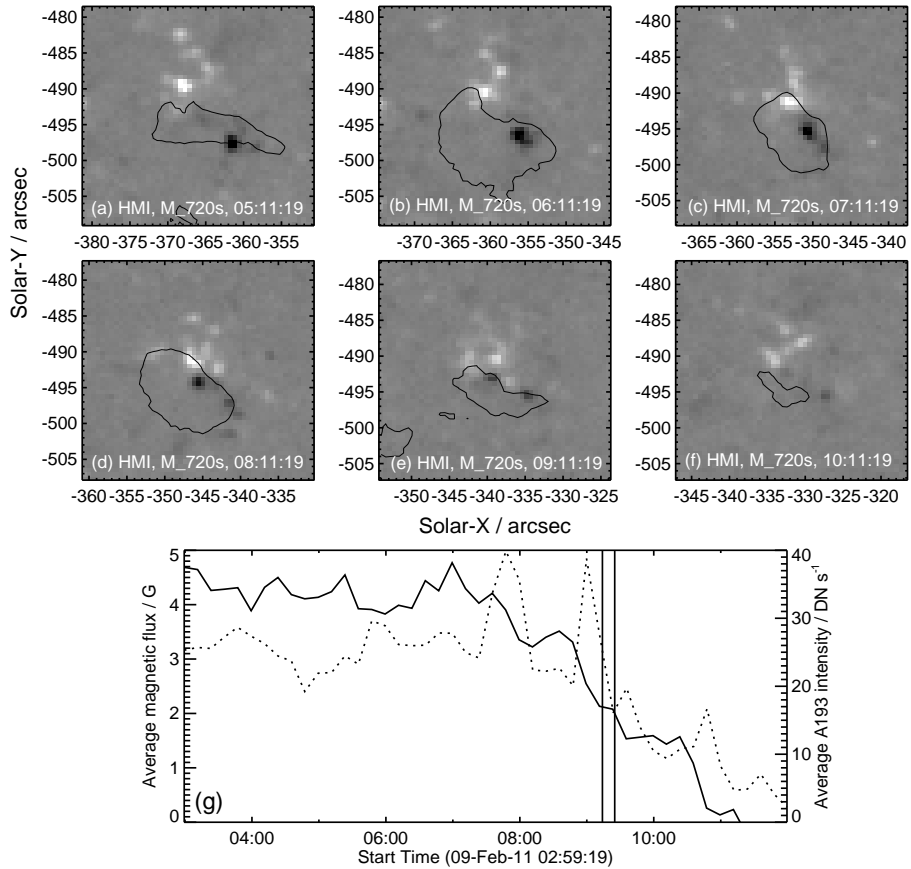}}
\caption{Panels a to f show HMI 720-second magnetograms at
  six different times. The magnetic flux has been saturated at
  $\pm$200~G in each image.  The contours show an \emph{A193} intensity level of
  100~DN~s$^{-1}$. Panel g shows the variation of unsigned
  magnetic flux (solid line) and \emph{A193} intensity (dotted line) as a
  function of time. Background levels have been subtracted for each quantity.}
\label{fig.hmi}
\end{figure}

The long-term evolution of the jet's bright point was studied with
HMI 720-second LOS magnetograms and 12-minute cadence \emph{A193} images
obtained over the period 03:00 to 12:00~UT.
The HMI 720-second magnetograms show positive and negative polarities
separated by about 10\as\ at 03:00~UT. They
come towards each other 
until they are separated by 2\,--\,3\as\ and
seem to annihilate each other between 09:00 and
10:00~UT. Figure~\ref{fig.hmi} shows six HMI images during the evolution
of the bright point and also the variation of the unsigned magnetic
flux with time. The magnetic flux was averaged over a 22\as\ $\times$
25\as\ region containing the strong magnetic polarities, and a background
level obtained by averaging over a wider 60\as\ $\times$ 60\as\
field of view was subtracted. The dotted line on
Figure~\ref{fig.hmi}g shows the variation of the \emph{A193} intensity
averaged over a 40\as\ $\times$ 37\as\ region containing the bright
point. A background value obtained from a nearby dark region within
the coronal hole was subtracted.

Figure~\ref{fig.hmi} shows that magnetic cancellation began at around
07:30~UT and continued until the average flux reached background
levels at 11:00~UT. The \emph{A193} intensity also falls over this period,
although there are three short episodes of intensity increases, one of
which immediately precedes the EIS raster scan of the event (indicated
by parallel vertical lines in
Figure~\ref{fig.hmi}g). The two other intensity increases did not give
rise to jet signatures in the \emph{A193} filter, and neither was observed by
EIS. 

\begin{figure} 
\centerline{\includegraphics[width=1.0\textwidth]{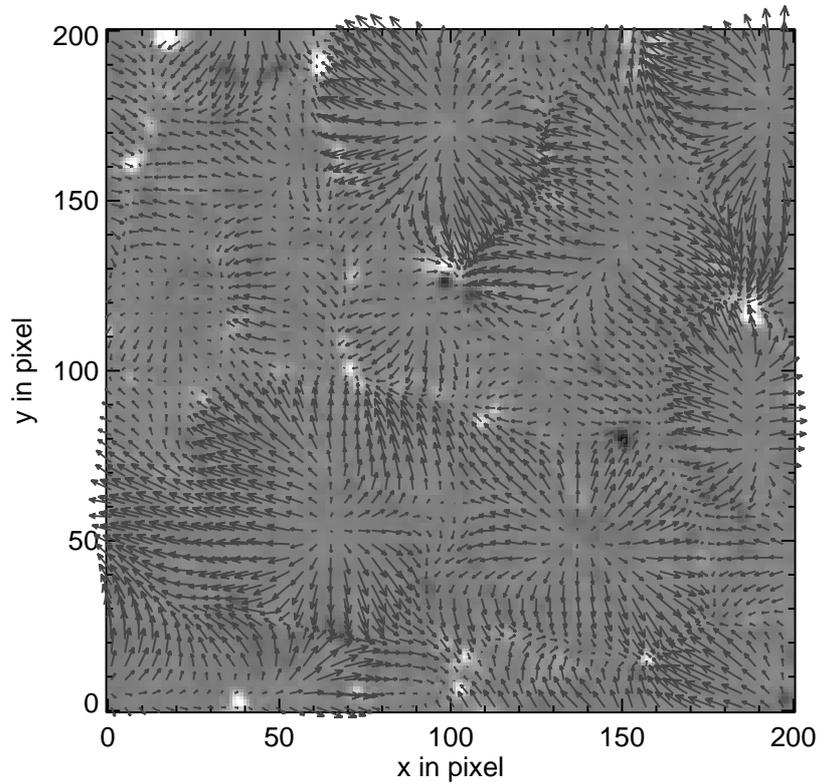}}
\caption{A LOS magnetogram averaged over the period 08:00 to 10:00~UT and
  saturated at a level of $\pm$100~G. The jet's bright point is
  located at
  (100,130), and a pixel corresponds to 0.6\as\ $\times$ 0.6\as. Arrows
  show the average plane-of-sky photospheric velocity field, 
  derived from cross-correlating HMI intensity images. The maximum
  flow-vector magnitude is 600~m~s$^{-1}$.}
\label{fig.lct}
\end{figure}

The photospheric magnetic field evolution seen in Figure~\ref{fig.hmi}
is dominated by flux convergence and
eventual cancellation. We use 45-second cadence HMI white-light images to
derive velocities perpendicular to the LOS using the Fourier Local
Correlation Tracking code (FLCT: \opencite{fisher08-flct}). 
The correlation is performed on pairs of images separated in time by
9~minutes as the typical lifetimes of granules are 10\,--\,15~minutes. In
addition, each image is convolved with a Gaussian function of
width 6.4\as\ as typical granule sizes are a few
arcseconds. (We note that the correlation was carried out with a range
of spatial scales, and the results are similar to those presented here.)
The flow field was calculated over a time period of two hours, from 08.00 to 10.00~UT,
and the field-of-view was centered on the location of the jet's bright point. The HMI images are
filtered to remove the intensity signatures of the solar $p$-modes (using
a phase velocity of 3~\kms).

Movie~2
shows the time-resolved flowfield at a 45-second cadence.  
Figure~\ref{fig.lct} shows the flow field averaged over two~hours to highlight the more
persistent flows. The background image of Figure~\ref{fig.lct} is the time averaged
magnetic field, saturated at $\pm$ 100~G, and the jet's underlying bipolar region
is at pixel position $x=100$, $y=130$. The flow field shows the characteristic
supergranular patterns with sizes of 20\,--\,30~Mm and velocities
up to a few hundred m~s$^{-1}$. At the jet location one can find the intersection
of four supergranules and the flows all converge at this point. The magnetic
flux elements found in the coronal hole are therefore swept together
leading to the observed flux convergence in the magnetogram sequence.
This is followed by flux cancellation as mixed polarities are involved
and the observed jet can be considered as the outflow region of the
reconnection of the flux system.
 
Section~\ref{sect.bp} focuses on the detailed evolution of the bright point
during the time period 08:48 to 09:15~UT when the jet was triggered,
and Section~\ref{sect.jet} presents an analysis of the ejected-jet
plasma.

\section{The Jet's Bright Point}\label{sect.bp}

\begin{figure} 
\centerline{\includegraphics[width=1.0\textwidth]{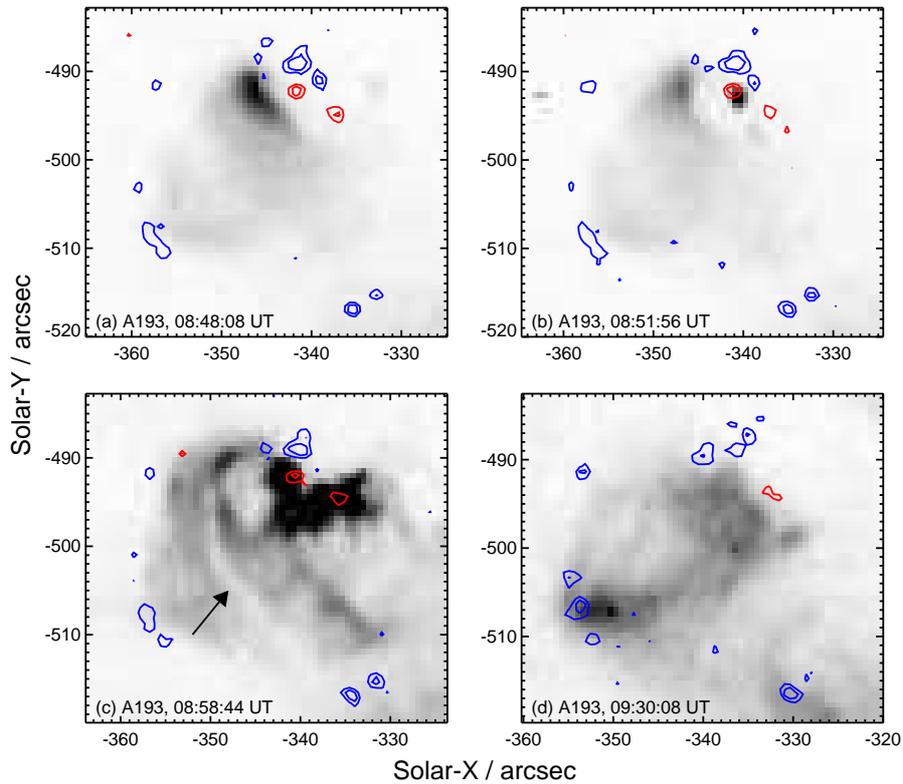}}
\caption{Four \emph{A193} images of the bright point that gives rise to the
  jet. A linear, reversed-intensity scaling is used for each image. The images
  in panels b and c have been saturated at 500 and 250~DN~s$^{-1}$,
  respectively. The actual maxima of the data are 784 and
  1064~DN~s$^{-1}$, respectively. Over-plotted on each image are
  contours showing the LOS magnetic-field strength, blue representing
  positive polarity and red negative polarity. The levels shown are 50
and 100~G.}
\label{fig.aia}
\end{figure}

In this section we describe the changes in the bright point during
08:48 to 09:15~UT, corresponding to the period
immediately prior and during the event. (For reference, this is the
second narrow intensity peak shown in
Figure~\ref{fig.hmi}g.)  Due to the rapid evolution of features within
the bright point we consider the 12-second
cadence AIA images, and Movie 3
and Movie 4
show \emph{A193} image sequences,
the former with a logarithmic intensity scaling to show better the
large scale loops of the bright point, and the latter with a
linear scaling to show better the intense brightenings at the bright
point base. 
Individual frames from the \emph{A193} sequence are shown
in Figure~\ref{fig.aia}.

In order to relate the evolution of the AIA images to
the magnetic field, it is necessary to co-align the AIA and HMI
images. For this purpose we consider images taken at around 08:51~UT,
which is when an intense, compact brightening is seen at the base of
the bright point. This is the first of a number of such brightenings, which
we refer to as ``kernels'' in this article. The kernel is present in the \emph{A1600} image at
08:51:06~UT, but is not present in the \emph{A1700} images around this
time. (We attribute this to the kernel emitting strongly in the
\ion{C}{iv} 1548 and 1550\,\AA\ emission lines that contribute to the \emph{A1600}
channel.) Now both the \emph{A1600} and \emph{A1700} images show bright points that
correspond to small magnetic features in the HMI images, and they can
be accurately aligned to HMI. We then cross-correlate the kernel seen
in the \emph{A1600} image with that seen in the \emph{A193} image, 
making the assumption that it is the same feature. This method allows
the HMI images to be aligned with the \emph{A193} images, and
Figure~\ref{fig.aia} shows four \emph{A193} images with magnetic field
contours overlaid.

Prior to the jet, at 08:48:08~UT (Figure~\ref{fig.aia}a), a small
loop-like shape was seen in the \emph{A193} image. If we take into account the
location of the bright point on the solar disk, then this loop-like
shape can be interpreted as joining the opposite polarities in the
magnetic field image. Note that the loop-like shape was spatially
offset from the magnetic features by a few arcseconds. We believe that
this was because the base of the loop-like shape is obscured by cool
material near the solar surface -- material that later erupted.

The first activity from the bright point was a single kernel that first
appeared at 08:50:20~UT within the dark area underneath
the small loop referred to earlier, and Figure~\ref{fig.aia}b shows
an image from 08:51:56~UT when the kernel is very intense -- at least a
factor three brighter than other parts of the bright point. The kernel
is about two pixels (900~km) wide in the $x$-direction and about five pixels (2200~km)
long in the $y$-direction. Note that this kernel and some others in the
sequence were partially removed by the AIA de-spiking routine (which
is intended to remove cosmic rays) and so the routine \textsf{AIA\_RESPIKE} had to be
applied to restore the correct intensity values. (The same procedure
also had to be applied to observations of flare kernels by \opencite{young13}.)

As can be seen from Figure~\ref{fig.aia}b,  the initial kernel appears
between two small, opposite polarity 
magnetic features, slightly closer to the
negative polarity feature (note that the coronal hole has an overall positive
polarity). The location of the kernel is considered accurate due to
the fact that the \emph{A1600} image shows both the kernel and the magnetic
features, as discussed earlier. The fact that the kernel is very close
to the magnetic features is evidence that the kernel is close to the
solar surface and not, say, occurring in the corona above the surface.

Inspection of Movie 3
shows that the kernel emits three small jets at
08:51:56, 08:52:56, and 08:54:08~UT. These extend away from the kernel in the southwest
direction and fade within about 10\as\ of the kernel's location. After
the third jet, the structure at the base of the bright point changes
significantly, with multiple kernels appearing along with expanding
loop-like shapes and jet structures. Figure~\ref{fig.aia}c shows the
appearance at 08:58:44~UT when multiple bright kernels are seen. Note
the kernels are arranged between the two negative polarity
magnetic fragments at the bright point base and are not directly connected
to the positive polarity fragments.

\begin{figure} 
\centerline{\includegraphics[width=1.0\textwidth]{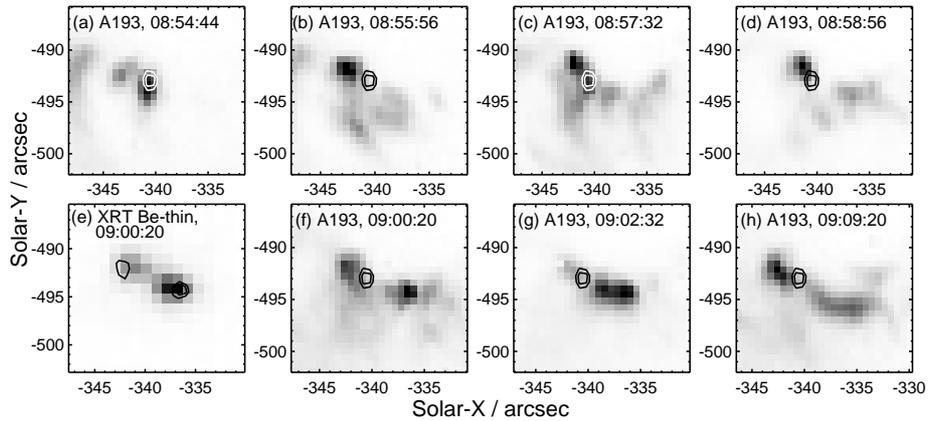}}
\caption{Panels a\,--\,d and f\,--\,h show \emph{A193} images of the kernels
  within the bright point at different times. On each image contours
  show the location of the kernel from 08:51:56~UT
  (Figure~\ref{fig.aia}). Panel e shows a XRT Be-thin image from
  09:00:20, the same time as the \emph{A193} image from panel f. The
  contours on the XRT image show the location of the \emph{A193} kernels from
panel f. In each image the color table has been reversed so that
black corresponds to a high intensity.}
\label{fig.kernels}
\end{figure}

The full evolution of the kernels in the \emph{A193} filter is best seen in
Movie 4, and Figure~\ref{fig.kernels} shows seven snapshots  of the
kernels at different times. Figure~\ref{fig.kernels} also shows one
XRT Be-thin image from 
09:00:20~UT (panel \ref{fig.kernels}e) that can be directly compared
to an \emph{A193} image from the 
same time (panel \ref{fig.kernels}f). On each of the \emph{A193}
images, contours from the \emph{A193} image at 
08:51:56~UT (Figure~\ref{fig.aia}b) are shown in order that the
kernels' positions can be related to the very first kernel. The
brightness of the kernels varies in time, so estimating lifetimes is
not straightforward. Emission is seen from the location of the first
kernel for about eight minutes, while emission is seen from the location of
the bright kernel at 08:55:56~UT for about 15~minutes. Significant brightness
variations can occur on a frame-by-frame basis (12-second cadence).

\begin{figure} 
\centerline{\includegraphics[width=1.0\textwidth]{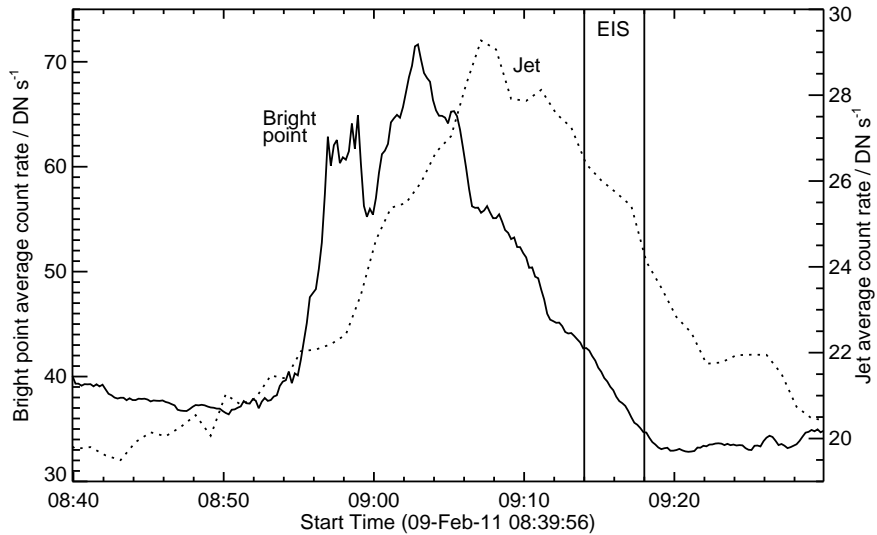}}
\caption{\emph{A193} light curves for the bright point (solid line) and a section
  of the extended jet emission (dotted line). The vertical solid lines
indicate the period when EIS scanned the jet. }
\label{fig.lc}
\end{figure}

A bright-point light curve was obtained by averaging the
  \emph{A193} emission over the individual image frames of Movie~3 and it reveals
one intensity peak 
between 08:54 and 09:00~UT and another between 09:00 and
09:08~UT (Figure~\ref{fig.lc}). Comparing with Movie 4, the ``spikiness'' of the first
peak is caused by rapid changes in the brightness of the kernels
between 08:56 and 08:59~UT. After a brief lull, the kernels brighten
again although the variation of intensity in time is more smooth and
also the kernels have a more East--West morphology  than
North--South, perhaps indicating that they are small loops rather than
jets in this period. We note also that the two XRT kernels at
09:00:20~UT (Figure~\ref{fig.kernels}e) are closer together than the
two \emph{A193} kernels at the same time (Figure~\ref{fig.kernels}f), which
may indicate that the kernels are actually two footpoints of a single,
small loop that is hotter at the apex. The variation of the  bright
point intensity with time  is not just driven by the kernels as the larger scale
loops expand and brighten, also. The contribution of the non-kernel
part of the bright point is greater during the second intensity
peak. The kernels have largely faded by around 09:10~UT, and by
09:30~UT, when the bright-point intensity has stabilized, the
morphology of the bright point has changed significantly
(Figure~\ref{fig.aia}d) with loops connecting to a magnetic
fragment in the southeast corner of the image being more intense than
the initial small loop (Figure~\ref{fig.aia}a).

Considering now the larger scale structure of the bright point,
Movie~3 reveals expanding loops that begin rising at around
08:54:00~UT, shortly 
after the first kernel appeared. An expanding ``dark
loop'' is apparent and is highlighted with an arrow in
Figure~\ref{fig.aia}c. From Movie 3 it can be seen that this dark
loop initially lay under the small loop shown in
Figure~\ref{fig.aia}a. For a loop to appear dark in the \emph{A193}
channel it must contain cool material (principally hydrogen and
helium) that is absorbing the coronal radiation. The expulsion of cool material during the event is
thus consistent with the blowout events studied by \inlinecite{moore10}. 
In addition to the expanding loops, the movie gives a general
impression of outflowing material from the bright point, for example
through filamentary structures that disconnect from the kernels and
flow outwards.

The temperatures reached during the jet evolution potentially provide
an important constraint on the physical mechanism behind the jet. 
For example, \inlinecite{moreno13} find temperatures as high as
7~MK in their model. The temperature within the extended jet feature
is discussed in the following section, and here we consider the bright
point temperature and, particularly, the temperatures reached in the
kernels. Temperatures are best determined with a spectroscopic
instrument but unfortunately EIS rastered across the bright point well
after the dominant activity within the bright point
(Figure~\ref{fig.lc}). Constructing an EIS spectrum of the bright
point shows that there is no emission from \ion{Fe}{xvi} and only a very
weak signal from \ion{Fe}{xv}, suggesting a peak temperature of
$<$~2~MK. 

The AIA image data potentially allow the kernel temperatures to be
measured, but Appendix~\ref{app.dem} demonstrates that differential
emission measure curves constructed from the data are not reliable.  Instead
we use the XRT filter data
as the ratio is sensitive to plasma
with temperatures
$>$~1~MK. The Ti-poly images are similar to the \emph{A193} images, although
there is a greater contrast between the kernels and the more-diffuse
bright point and jet emission. The Be-thin filter images show only the
kernels, immediately demonstrating that they have a higher temperature
than the bright point or jet (since the filter is sensitive to higher
temperatures than Ti-poly). The ratio of the filters can
yield a temperature, and we take three of the kernels recorded between 08:57 and
09:03~UT, sum their
intensities, and compute the intensity ratios. The Be-thin/Ti-poly
intensity ratios are 0.05, 0.41, and 
0.14 which convert to temperatures of 1.5\,--\,3.0~MK (see, \textit{e.g.},
Section~3 of \opencite{reeves12}). We caution that the Be-thin and
Ti-poly images were taken 30 seconds apart and the AIA data show that the
kernels can evolve on 
this time-scale, but we do not expect actual temperatures to be
significantly different from the stated values.



In summary, the XRT filter-ratio data provide a strong constraint on
the maximum temperature in the kernels with values of $\le
3$~MK. If there is very hot plasma (10~MK) then the amount is not
significant enough to affect the XRT filter ratio measurements.

\section{The Extended Jet Emission}\label{sect.jet}

The jet extending away from the bright point is not a narrow columnar
structure but instead is rather broad (a ``curtain'' in the notation
of \opencite{moore10}). The precise width of the jet is difficult to
estimate due to contamination by quiet Sun plasma on the west side of
the jet, but we estimate that it is approximately 40\as\ (30~Mm). The
width increases as the event progresses. In the \emph{A193} images 
the jet extends about 120\as\ (87~Mm) from the bright point before
becoming too faint to see. The jet is clearly
not directed in a radial direction, but  the east side of the
jet is seen to curve towards the Southeast suggesting that
the plasma is tending towards a radial 
direction with height. There is also evidence of
fine-scale 
filamentary structure along the jet's axis in Movie 1.

The jet is best seen in the \emph{A193} filter, but is also seen in \emph{A171}
(dominated by \ion{Fe}{ix}) and
very weakly in \emph{A131} (which we take to be \ion{Fe}{viii} emission,
\opencite{odwyer10}). There is no jet emission in \emph{A335}, implying the
jet is not hot enough to produce \ion{Fe}{xvi}, an ion principally
formed over the temperature range 2.2\,--\,4.5~MK. As \ion{Fe}{xvi}
335.4\,\AA\ is a strong transition, this provides a strong constraint for
the temperature to be $\le$~2~MK independent of the uncertainties of
AIA differential emission measure analysis (Appendix~\ref{app.dem}). A movie formed in the 
\emph{A304} channel does suggest plasma extending above the bright point,
consistent with the jet, but it is not clearly seen due to the
background coronal hole emission. Movies in these filters are
available at
\href{http://pyoung.org/jets/hop177/jet\_27.html}{\textsf{pyoung.org/jets/hop177/jet\_27.html}}.

The velocity of the jet in the plane of the sky was estimated by
considering \emph{A193} light curves at two different heights. The light curves
were derived by averaging the \emph{A193} intensity over $5\times 3$ blocks
of pixels. By adjusting the
position of the light curve for the greater height to match that for
the lower height, the travel time for the jet front could be
estimated. Due to the lateral broadening of the jet with time, care
was taken to select two locations that brightened during the initial
rise of the jet, before the lateral broadening began. We find a
projected velocity of 203~\kms\ at a projected height of 71~Mm above
the kernels' location. We estimate an uncertainty of 25\,\%\ in this measurement
due to the noise in the light curves.

Considering a larger portion of the jet, Figure~\ref{fig.lc} compares
the \emph{A193} light curve 
 with that of the bright point (which was
discussed in Section~\ref{sect.bp}). The jet light curve was obtained
for a 15\as\ $\times$ 15\as\ region that is within the blue-shifted
column of Figure~\ref{fig.eis}. The region was chosen to be in the
coronal hole and not to contain any bright-point emission. Although
the jet emission peaks later than the bright-point emission, it began
rising at a similar time to the bright point, suggesting
that plasma was ejected from very early in the event. Note that in
Section~\ref{sect.bp} a small jet from the inital kernel was identified
as early as 08:51:56~UT. The vertical
lines indicate the time period when EIS scanned the jet, and it can be
seen that the emission had already decayed significantly at this time.

\begin{figure} 
\centerline{\includegraphics[width=1.0\textwidth]{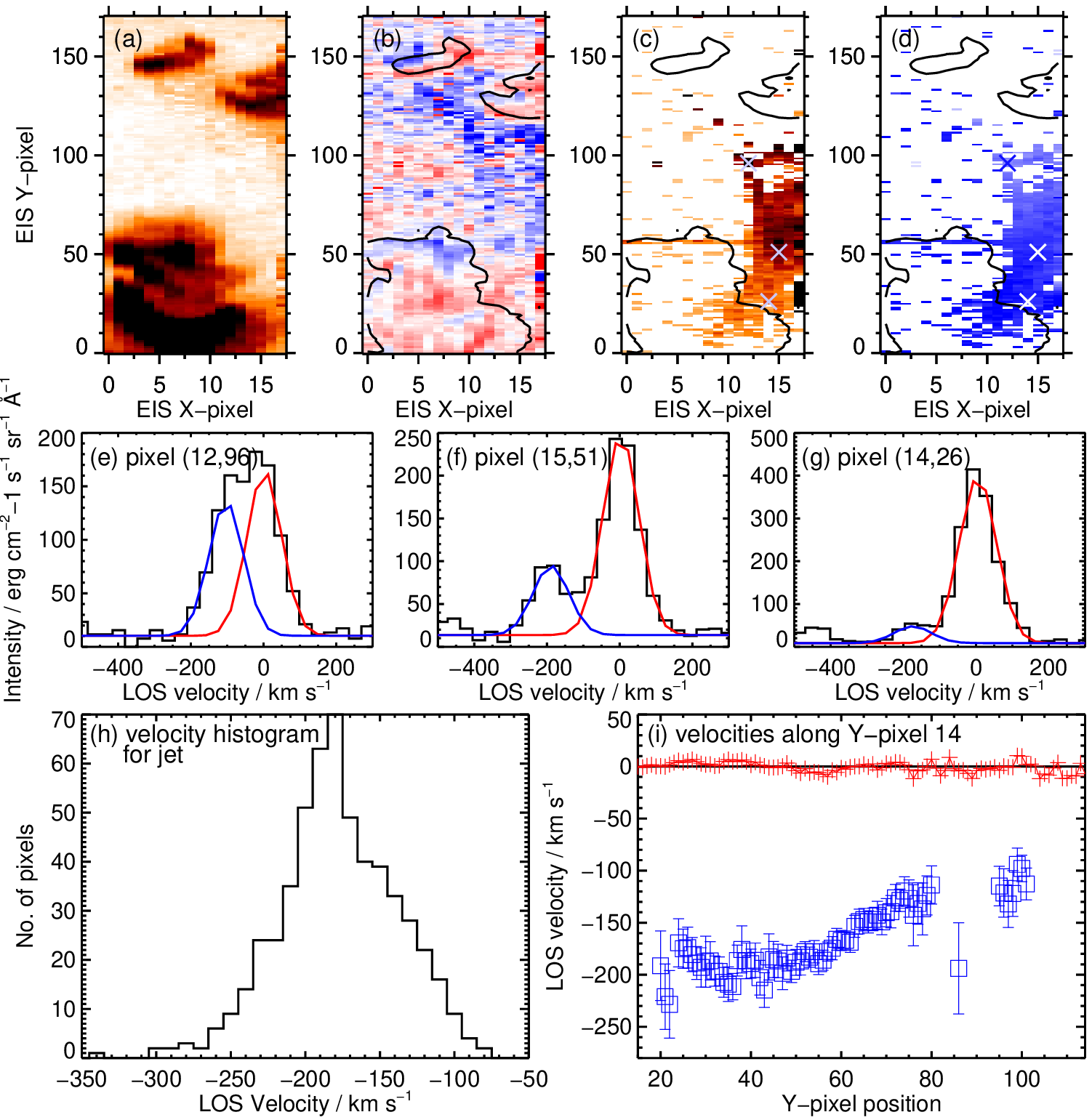}}
\caption{Results from the two Gaussian fit to the EIS \ion{Fe}{xii}
  195.12\,\AA\ emission line. Panels a and c show intensity maps for
the two components with a reverse intensity scaling applied, and panel
a is saturated at 100~\ecss. Panels b and d show velocity
maps. Contours on panels b\,--\,d show an intensity level of 50~\ecss\
from panel a. Crosses
on panels c and d indicate the spatial pixels for which the spectra
shown in panels e--g were extracted. The red and blue
curves in panels e--g show the two Gaussian components. Panel h shows the
velocity histogram for the high-velocity jet component extracted from
$x$-pixels 9 to 17 and $y$-pixels 0 to 109. Panel g shows a slice
through the two velocity maps (panels b and d) at $y$-pixel 14.}
\label{fig.eis-2g}
\end{figure}

Inspection of images formed in different EIS emission lines suggests
that the jet had a rather narrow temperature structure, consistent with the
AIA images. An image formed in \ion{Fe}{xi} 188.2\,\AA\ is similar to
that of \ion{Fe}{xii} 195.1\,\AA\, but \ion{Fe}{xiii} 202.0\,\AA\  shows a
much weaker signature. \ion{Fe}{ix} 197.9\,\AA\  and \ion{Fe}{x}
184.5\,\AA\  images show emission in the lower part of the jet, \textit{i.e.}
$y$-values of $-520$ to $-560$ in Figure~\ref{fig.eis}. \ion{He}{ii}
256.3\,\AA\  and \ion{O}{v} 192.9\,\AA\  show some emission just above the
bright point, but no obvious emission within the main body of the jet.

The \ion{Fe}{xii} 195.12\,\AA\  emission line profiles in the
jet clearly demonstrate a two component structure, which we interpret
as a superposition of the jet plasma and the background coronal hole
plasma. For this reason a two-Gaussian fit was applied to the line
profile, and Figure~\ref{fig.eis-2g} shows the results. The intensity and
velocity maps for the ``background'' component are shown in panels \ref{fig.eis-2g}a
and \ref{fig.eis-2g}b. The bright point that is the source of the jet is seen in the
top-left corner of the intensity image. (A much larger and more intense bright
point, not related to the jet, is seen at the bottom of the image.) The velocity image for the
background component is scaled between $-15$~\kms\ and $+15$~\kms\ and
weak blueshifts are seen above the bright point. Panels \ref{fig.eis-2g}c and \ref{fig.eis-2g}d
show the intensity and velocity maps for the ``jet'' Gaussian
component. Where reliable two Gaussian fits were not possible, these
pixels are white in the two images. It can be seen that the second
component is  present mainly in the thick columnar structure
previously highlighted in Figure~\ref{fig.eis}. This demonstrates that
the 
enhanced line width seen in the single Gaussian fit is actually due to
the presence of an extra Gaussian component. 

Crosses on panels \ref{fig.eis-2g}c and \ref{fig.eis-2g}d indicate the spatial locations for which
the line profiles in panels \ref{fig.eis-2g}e\,--\,g were extracted. For locations in
the column that are closer to the bright point, the jet velocity is
smaller and so the line profile has the appearance of a single broad
line (panel \ref{fig.eis-2g}e). For locations further from the bright point, the two
Gaussian components are clearly separated (panels \ref{fig.eis-2g}f and \ref{fig.eis-2g}g). A
histogram showing all of the jet velocities is shown in panel \ref{fig.eis-2g}h. The
distribution has a mean of $-173$~\kms\ and a standard deviation of
42~\kms. Panel \ref{fig.eis-2g}i shows a slice through the two velocity images
(panels \ref{fig.eis-2g}b and \ref{fig.eis-2g}d) at $x$-pixel 14. This demonstrates that the background
component (red) is close to rest velocity while the jet component (blue)
increases with distance from the bright point from about $-100$~\kms\
to about $-200$~\kms.

This latter finding is important as it implies that the plasma is
being accelerated as it travels away from the bright point. Previous
authors have noted that jet velocities measured from image sequences
are often not sufficient for 
the plasma to escape from the Sun \cite{culhane07}. Spectroscopic
velocity measurements are a more accurate reflection of the true
plasma velocity, and the increasing velocity with height found here
suggests that jet plasma could escape the Sun even if the velocities
reported at low heights are below the escape velocity. The theoretical
model of \inlinecite{pariat09} demonstrated that nonlinear Alfv\'en waves
can escape along open field lines following magnetic reconnection of
twisted, closed magnetic field within the bright point. Such waves
could be responsible for accelerating the plasma in the jet.

At the height for which the projected jet velocity was derived from
the \emph{A193} data, the EIS LOS velocity is $\approx$~100~\kms, so the
projected velocity is around a factor two larger than the LOS
velocity. The cosine factor at the location of the bright point is
0.77 so for a radial jet one would expect the LOS velocity to be
larger than the projected velocity. Therefore, at least at relatively low
heights, the jet has a somewhat ``flat'' trajectory. We caution,
however, that the projected velocity is determined by the initial
intensity front of the jet at $\approx$~09:02~UT while the EIS
LOS velocity measurement reflects measurements made during the period
09:14\,--\,09:18~UT.

Density and temperature can be estimated in the jet by considering
additional lines to 195.12\,\AA; however for these lines it is not
possible to make reliable measurements in individual spatial pixels
so it is necessary to bin multiple pixels. We chose a 4$\times$21
block of pixels within the jet and used the IDL routine
\textsf{EIS\_MASK\_SPECTRUM} to create the averaged spectrum; this routine
corrects for spatial offsets between different wavelengths.
\ion{Fe}{xii} 186.9\,\AA/195.12\,\AA\ is a good density diagnostic
(\opencite{young07-eis} and \opencite{young09-dens}), and the
186.9\,\AA\ profile shows a weak, blueshifted component that can be
identified with the jet. Using atomic data from version 7.1 of the
CHIANTI database (\opencite{dere97}; \opencite{chianti71}) yields a
density of $\log\,(N_{\rm e}[{\rm
  cm}^{-3}])=8.44^{+0.29}_{-0.39}$ for the weak blue-shifted component
corresponding to the jet plasma. Selecting a dark region
  directly to
the East of the jet, we find a density of $\log\,(N_{\rm e}[{\rm
  cm}^{-3}])=8.67\pm 0.05$ thus the jet density is consistent with this
value, \textit{i.e.} the jet does not show any significant density increase
over the background coronal hole.
The average intensity measured in the jet is
5.2~\ecss\ so converting to a column depth yields a value of 1300~km,
much smaller than the observed width of the jet. This suggests that a
curtain is an apt description of the jet, it being much thinner in the
line-of-sight direction than in the plane of the sky.

An additional  jet component to the background component can only be
clearly identified for the two \ion{Fe}{xii} lines discussed in the
previous paragraph and \ion{Fe}{xi} 188.22\,\AA\ and \ion{Fe}{xiii}
202.04\,\AA. Again using atomic data from CHIANTI, the \ion{Fe}{xiii}
202.04\,\AA/\ion{Fe}{xii} 195.12\,\AA\ and \ion{Fe}{xii}
195.12\,\AA/\ion{Fe}{xi}  188.22\,\AA\ ratios yield isothermal
temperatures of $\log\,T=6.16$ and 6.12, respectively, so we can say
that the ejected plasma has a temperature of $\approx$~1.4~MK.

\section{Summary}\label{sect.summary}

A jet observed in the south coronal hole at 9 February 2011 09:00~UT is identified as a blowout
jet based on the disruption of the base arch of the bright point and
the ejection of cool plasma during the event.
The combination of high-cadence coronal imaging
from AIA and XRT, magnetograms from HMI and coronal spectroscopy from
EIS give important new insights into the processes of blowout
jets. The key results are:

\begin{itemize}
\item The jet has a lifetime of $\approx$ 30~minutes and occurs during a
  magnetic cancellation event within a 
  small coronal bright point. The cancellation occurs over a period of
  four hours and ultimately leads to the
  disappearance of the bright point.
\item Photospheric flows derived through local correlation tracking
  show that the bright point lies in a region of converging flow. 
\item The bright point shows a number of bright kernels typically of
  size 1000~km. Lifetimes range from 1 to 15~minutes and brightness
  variations occur on scales of tens of seconds. The kernels are spatially located close
  to the minority polarity features of the bright point, and they have
  temperatures of 1\,--\,3~MK.
\item Jet emission appears immediately after activity begins in the
  bright point, although the peak jet emission occurs five minutes after the
  peak bright point emission. The jet continues to be enhanced over
  the coronal background ten minutes after the bright point becomes
  quiescent. 
\item A loop of cool plasma (a mini-filament?) is ejected during the
  event and there is evidence that the loop was present in the bright
  point prior to the jet.
\item The jet plasma shows outflowing plasma with LOS speeds of 
  100\,--\,250~\kms\ (with an average value of 173~\kms), a density
  of $2.8\times 10^8$~cm${-3}$, and a temperature of 1.4~MK.
\item The LOS velocity increases along the jet, from 100~\kms\ to
  250~\kms\ suggesting that the plasma is being accelerated within the jet,
  not just at the bright point.
\end{itemize}

%

%
\begin{acknowledgment}
The authors acknowledge funding from National Science Foundation grant
AGS-1159353. Valuable comments from the anonymous referee and
E.~Pariat are acknowledged. SDO is a mission for NASA's Living With a
Star program. 
Data are provided courtesy of NASA/SDO and the AIA and HMI science teams.
\hinode\ is a Japanese mission developed and launched by 
ISAS/JAXA, with NAOJ as domestic partner and NASA and
STFC (UK) as international partners. It is operated by
these agencies in co-operation with ESA and NSC (Norway).
\end{acknowledgment}

%
%
\bibliographystyle{spr-mp-sola}
\bibliography{myrefs_solphys}  



\appendix

\section{AIA DEM Analysis of Bright Point Kernels}\label{app.dem}

A differential emission measure analysis of two of the bright point
kernels was attempted using the AIA filter data, but the results are
not considered accurate due to the large emission measure found at
$\approx$ $10^7$~K, which is inconsistent with the XRT filter ratio
results. Here we record the method used to derive the AIA DEM curves.

\begin{figure} 
\centerline{\includegraphics[width=1.0\textwidth]{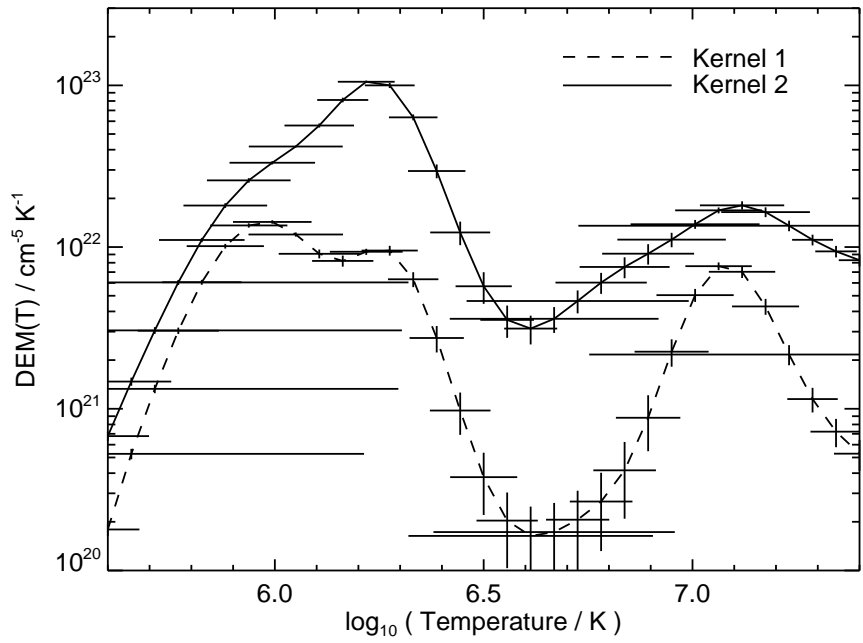}}
\caption{Differential emission measure curves for two kernels within
  the jet bright point, derived from AIA filter intensities.}
\label{fig.aia-dem}
\end{figure}

The multiple filters of AIA give access to a wide range of
temperatures; however for most of the channels the temperature
response functions show multiple peaks due to emission lines formed at
different temperatures contributing to the
passband \cite{boerner12}. \inlinecite{hannah12} presented a differential emission
measure method that can be applied to AIA data, and this was applied
to the initial kernel using images obtained between 08:51:49 and
08:51:59~UT, and also a bright elongated feature observed between
09:02:26 and 09:02:37~UT (Figure~\ref{fig.kernels}g) that probably
consists of two or three 
kernels. We refer to these two features as ``kernel 1'' and ``kernel
2'', respectively. The kernel intensities for the six coronal filters,
\emph{A94}, \emph{A131}, 
\emph{A171}, \emph{A193}, \emph{A211}, and \emph{A335}, were measured
and a background intensity 
was subtracted for each of the \emph{A131},
\emph{A171}, \emph{A193}, and \emph{A211} filters (the background level
for the remaining 
filters was negligible). The values are given in
Table~\ref{tbl.aia-ints}. The \inlinecite{hannah12} method yielded a
continuous DEM curve between $\log\,T=5.6$ and 7.4 for each kernel and
they are shown in Figure~\ref{fig.aia-dem}. (We note that the
\textsf{/noblend}, \textsf{/evenorm}, and \textsf{/chiantifix} keywords were applied when
retrieving the AIA response functions.) Both curves show emission at low
temperatures ($\log\,T=5.9$ to 6.4) with a further peak at
$\log\,T=7.1$, suggesting there is a significant amount of very hot
plasma in the kernels. However, this result is at odds with the XRT
observations, since both 
the Ti-poly and Be-thin filters are much more sensitive at $10^7$~K
than $10^6$~K. Folding the XRT response curves with the derived DEMs
yields predicted Be-thin/Ti-poly ratios of 0.68 and 0.66 for the two
kernels, much higher than the measured ratios during the XRT sequence.
The reason for the spurious DEM curves from AIA is likely due to
contributions of cool lines to the \emph{A94} and \emph{A131} channels that are
currently unaccounted for in atomic models.

\begin{table}
\caption{Kernel intensity measurements.
}
\label{tbl.aia-ints}
\begin{tabular}{ccccccc}     
  \hline                   
 & \multicolumn{6}{c}{Intensity / DN s$^{-1}$} \\
\cline{2-7}
Kernel & 94 & 131 & 171 & 193 & 211 & 335 \\
  \hline
1 & 64 & 769 & 8791 & 4995 & 1420 & 89 \\
2 & 8 & 50 & 703 & 929 & 278 & 14 \\
  \hline
\end{tabular}
\end{table}

\end{article} 
\end{document}